\documentclass[journal]{IEEEtran}
\usepackage{graphicx,graphicx,epsfig}

\ifCLASSINFOpdf
\else
\fi

\begin{document}

\title{Experimental demonstration of associative memory with memristive neural networks}
%
%
%

\author{Yuriy~V.~Pershin and Massimiliano~Di~Ventra
\thanks{Yu. V. Pershin is with the Department of Physics
and Astronomy and USC Nanocenter, University of South Carolina,
Columbia, SC, 29208 \newline e-mail: pershin@physics.sc.edu.}
\thanks{M. Di Ventra is with the Department
of Physics, University of California, San Diego, La Jolla,
California 92093-0319 \newline e-mail: diventra@physics.ucsd.edu.}
\thanks{Manuscript received November XX, 2009; revised January YY, 2010.}}

\maketitle

\begin{abstract}
Synapses are essential elements for computation and information
storage in both real and artificial neural systems. An artificial
synapse needs to remember its past dynamical history, store a
continuous set of states, and be "plastic" according to the
pre-synaptic and post-synaptic neuronal activity. Here we show that
all this can be accomplished by a memory-resistor ({\it memristor}
for short). In particular, by using simple and inexpensive
off-the-shelf components we have built a memristor emulator which
realizes all required synaptic properties. Most importantly, we have
demonstrated experimentally the formation of associative memory in a
simple neural network consisting of three electronic neurons
connected by two memristor-emulator synapses. This experimental
demonstration opens up new possibilities in the understanding of
neural processes using memory devices, an important step forward to
reproduce complex learning, adaptive and spontaneous behavior with
electronic neural networks.
\end{abstract}


%
\IEEEpeerreviewmaketitle

\section{Introduction}

\IEEEPARstart{W}{hen} someone mentions the name of a known person
we immediately recall her face and possibly many other traits.
This is because we possess the so-called associative memory - the
ability to correlate different memories to the same fact or event
\cite{litr1}. This fundamental property is not just limited to
humans but it is shared by many species in the animal kingdom.
Arguably the most famous example of this are experiments conducted
on dogs by Pavlov \cite{litr2} whereby salivation of the dog's
mouth is first set by the sight of food. Then, if the sight of
food is accompanied by a sound (e.g., the tone of a bell) over a
certain period of time, the dog learns to associate the sound to
the food, and salivation can be triggered by the sound alone,
without the intervention of vision.

\begin{figure}[b]
\centering
\includegraphics[width=8.5cm]{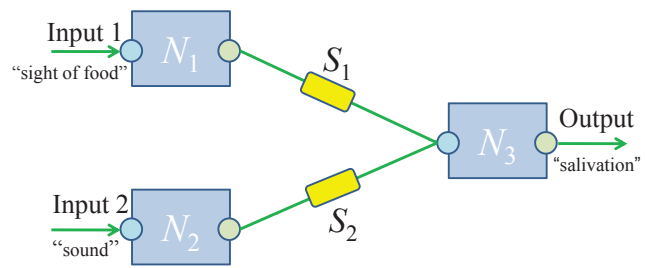}
 \caption{Artificial neural network for associative memory. Real neurons
and their networks are very complex systems whose behavior is not
yet fully understood. However, some simple brain functions can be
elucidated studying significantly simplified structures. Here, we
consider three neurons ($N_1$, $N_2$ and $N_3$) coupled by two
memristive synapses ($S_1$ and $S_2$). The output signal is
determined by input signals and strengths of synaptic connections
which can be modified when learning takes place. \label{fig1}}
\end{figure}

Since associative memory can be induced in animals and we, humans,
use it extensively in our daily lives, the network of neurons in our
brains must execute it very easily. It is then natural to think that
such behavior can be reproduced in artificial neural networks as
well - a first important step in obtaining artificial intelligence.
The idea is indeed not novel and models of neural networks have been
suggested over the years that could theoretically perform such
function \cite{hopf,litr3,litr5,litr4}. However, their experimental
realization, especially in the electronic domain, has remained
somewhat difficult. The reason is that an electronic circuit that
simulates a neural network capable of associative memory needs two
important components: neurons and synapses, namely connections
between neurons. Ideally, both components should be of nanoscale
dimensions and consume/dissipate little energy so that a scale-up of such
circuit to the number density of a typical human brain (consisting
of about 10$^{10}$ synapses/cm$^2$) could be feasible. While one
could envision an electronic version of the first component
relatively easily, an electronic synapse is not so straightforward
to make. The reason is that the latter needs to be flexible
("plastic") according to the type of signal it receives, its
strength has to depend on the dynamical history of the system, and
it needs to store a continuous set of values (analog element).

In the past, several approaches with different levels of
abstraction were used in order to implement electronic analogues
of synapses \cite{lewis}. For instance, one of the first ideas
involved the use of three-terminal electrochemical elements
controlled by electroplating~\cite{widrow}. While some of these
approaches do not involve synaptic plasticity at all
\cite{no_plast1,no_plast2,no_plast3,no_plast4}, the latter is
generally implemented using a digital (or a combination of analog
and digital) hardware
\cite{plast1,plast2,plast3,plast4,plast5,plast6,plast7}. The
common feature of synaptic plasticity realizations is the
involvement of many different circuit elements (such as
transistors) and, therefore, occupation of a significant amount of
space on a VLSI chip. Thus, the amount of electronic synapses in
present VLSI implementations is much lower than the amount of
synapses relevant to actual biological systems. Novel, radically
different approaches to resolve this issue would be thus
desirable.

A recently demonstrated resistor with memory (memristor for short
\cite{litr6,litr7}) based on TiO$_2$ thin films
\cite{litr9,litr10} offers a promising realization of a synapse
whose size can be as small as 30$\times$30$\times$2 nm$^3$. Using
TiO$_2$ memristors, a fabrication of neuromorphic chips with a synapse density
close to that of the human brain may become possible. Memristors
belong to the larger class of memory-circuit elements (which
includes also memcapacitors and meminductors) \cite{litr81,litr8},
namely circuit elements whose response depends on the whole
dynamical history of the system. Memristors can be realized in
many ways, ranging from oxide thin films
\cite{litr9,litr10,litr11} to spin memristive systems
\cite{litr12}. However, all these realizations are limited to the
specific material or physical property responsible for memory, and
as such they do not easily allow for tuning of the parameters
necessary to implement the different functionalities of electronic
neural networks.

In the present paper, we describe a flexible platform allowing for
simulation of different types of memristors, and experimentally show
that a memristor could indeed function as a synapse. We have
developed electronic versions of neurons and synapses whose behavior
can be easily tuned to the functions found in biological neural
cells. Of equal importance, the electronic neurons and synapses were
fabricated using inexpensive off-the-shelf electronic components
resulting in few dollars cost for each element, and therefore can be
realized in any electronic laboratory. Clearly, we do not expect
that with such elements one can scale up the resulting electronic
neural networks to the actual brain density. However, due to their
simplicity reasonably complex neural networks can be constructed
from the two elemental blocks developed here and we thus expect
several functionalities could be realized and studied.

For the purpose of this paper we have built the neural network shown
in Fig. \ref{fig1}. We have then shown that such circuit is capable
of associative memory. In this network, two input neurons are
connected with an output neuron  by means of synapses. As an example
of the functionality that this network can provide, we can think
about the animal memory we have described above \cite{litr2} in
which the first input neuron (presumably located in the visual
cortex) activates under a specific visual event, such as "sight of
food", and the second input neuron (presumably located in the
auditory cortex) activates under an external auditory event, such as
a particular "sound". Depending on previous training, each of these
events can trigger "salivation" (firing of the third output neuron).
If, at a certain moment of time, only the "sight of food" leads to
"salivation," and subsequently the circuit is subjected to both
input events, then, after a sufficient number of simultaneous input
events the circuit starts associating the "sound" with the "sight of
food", and eventually begins to "salivate" upon the activation of
the "sound" only. This process of learning is a realization of the
famous Hebbian rule stating, in a simplified form, that "neurons
that fire together, wire together".

\section{Results and Discussion}
\subsection{Electronic neuron}

Biological neurons deal with two types of electrical signals:
receptor (or synaptic) potentials and action potentials. A
stimulus to a receptor causes receptor potentials whose amplitude
is determined by the stimulus strength. When a receptor potential
exceeding a threshold value reaches a neuron, the latter starts
emitting action potential pulses, whose amplitude is constant but
their frequency depends on the stimulus strength. The action
potentials are mainly generated along axons in the forward
direction. However, there is also a back-propagating part of the
signal \cite{litr13,litr14} which is now believed to be
responsible for synaptic modifications (or learning)
\cite{litr14,litr15}.

\begin{figure}[t]
\centerline{\includegraphics[width=6cm]{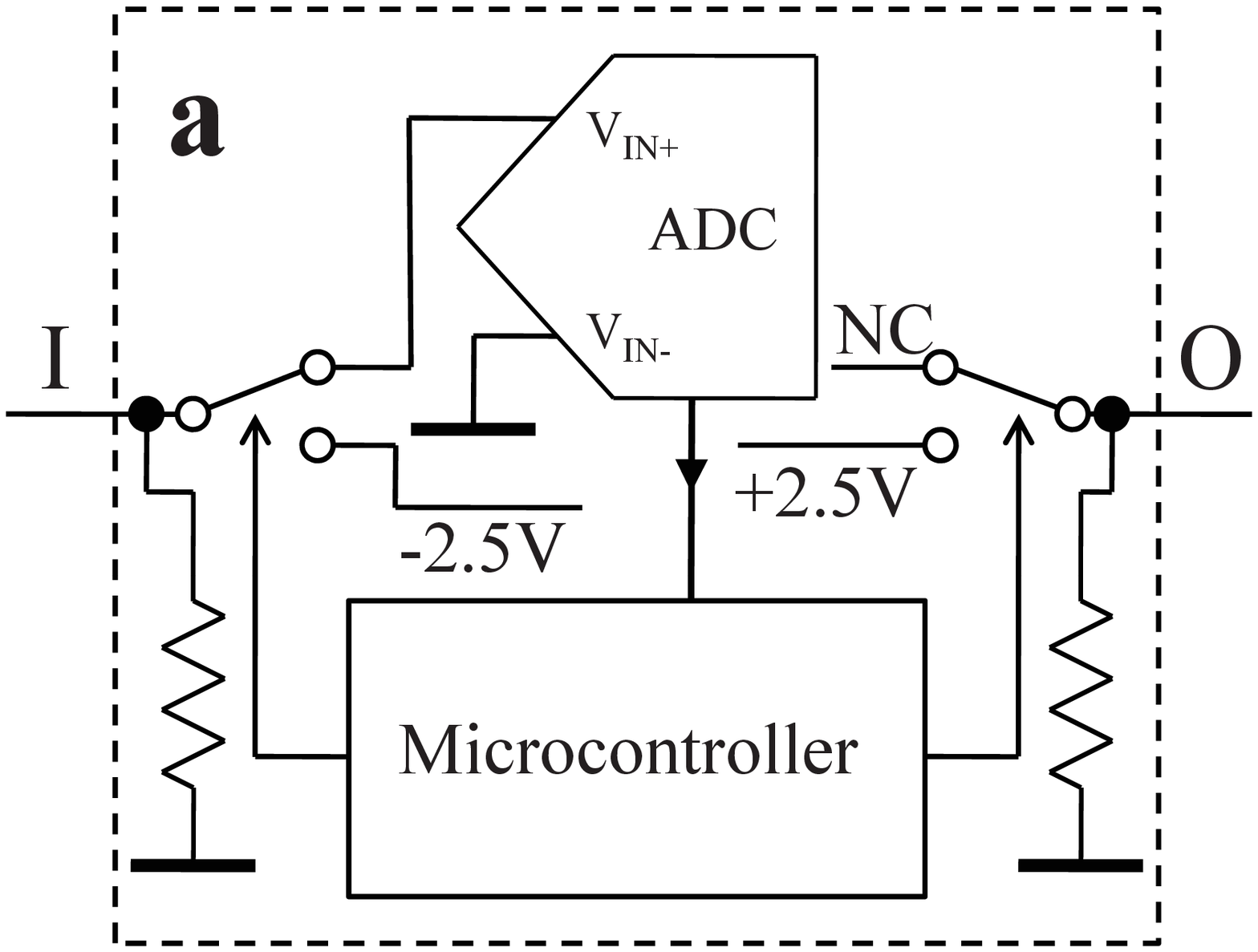}}
\centerline{\includegraphics[width=8cm]{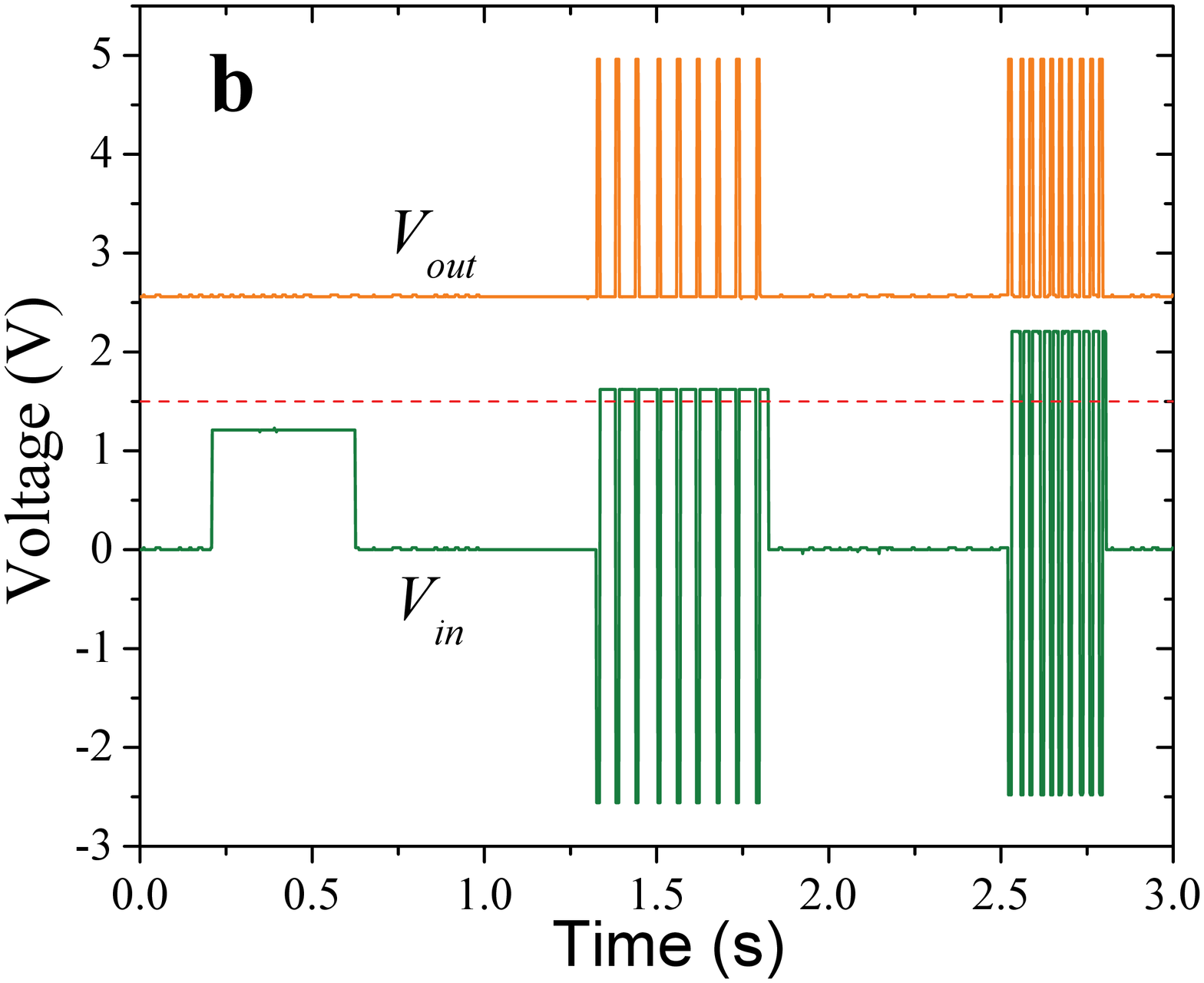}}
 \caption{Electronic neuron. {\bf a}, Main components of the electronic neuron
proposed here are an analog-to-digital converter (ADC) and a
microcontroller. NC means "not connected". If the voltage value
$V_{in}$ on the input terminal (I) exceeds a threshold voltage $V_T$
(in our experiments $V_T= 1.5$V), the microcontroller connects the
input pin to -2.5V and output pin (O) to 2.5V for 10ms, thus
sending forward and backward pulses. After that, it waits for a
certain amount of time $\delta t$ and everything repeats again. If
$V_{in} < V_T$, then the microcontroller just continuously samples
$V_{in}$. The waiting time was selected as $\delta t=\tau-\gamma
\cdot \left(V_{in}-V_T \right)+\lambda\left(\eta-0.5 \right)$,
where $\tau=60$ms, $\gamma = 50$ms/V, $\lambda =10$ms and $\eta$
is a random number between 0 and 1. In our experimental
realization, we used microcontroller dsPIC30F2011 from Microchip
with internal 12bits ADC and a possibility of pin multiplexing, so
that the only additional elements were two 10kÙ resistors. Actual
measurements were done using 5V voltage span, with further
assignment of the middle level as zero. The value of resistors
shown in {\bf a} is 10k$\Omega$. {\bf b}, Response of the
electronic neuron on input voltages of different magnitude, the
red dashed line is the threshold voltage and is only a guide to
the eye. When $V_{in}<V_T$, no "firing" occurs. When $V_{in}>V_T$,
the electronic neuron sends pulses, with the average pulse
separation decreasing with increasing $V_{in}$. $V_{out}$ is
shifted for clarity. \label{fig2}}
\end{figure}

We have implemented the above behaviour in our electronic scheme
as shown in Fig. \ref{fig2}a using an analog-to-digital converter
and a microcontroller. The input voltage is constantly monitored
and once it exceeds a threshold voltage, both forward and backward
pulses are generated whose amplitude is constant (we set it here
for convenience at 2.5V), but pulse separation varies according to
the amplitude of the input signal. In Fig. \ref{fig2}b we show the
response of the electronic neuron when three resistors of
different values are subsequently connected between the input of
the neuron and 2.5V. When the resulting voltage (determined by the
external resistor and internal resistor connected to the ground in
Fig. \ref{fig2}a) is below the threshold voltage ($t<1$s in Fig.
\ref{fig2}b), no "firing" occurs (no change in the output
voltage). When the input voltage exceeds the threshold ($t>1$s in
Fig. \ref{fig2}b), the electronic neuron sends forward- and
back-propagating pulses. The pulse separation decreases with
increase of the input voltage amplitude as it is evident in Fig.
\ref{fig2}b.

\subsection{Electronic synapse}

As electronic synapse we have built a {\it memristor emulator},
namely an electronic scheme which simulates the behaviour of any
memory-resistor. In fact, our memristor emulator can reproduce the
behaviour of any voltage- or current-controlled memristive system.
The latter is described by the following relations

\begin{eqnarray}
y\left( t \right) =g\left( x,y,t \right) u\left( t\right),
\label{eq1}
\\
\dot x=f\left( x,u,t\right), \label{eq2}
\end{eqnarray}
where $y(t)$ and $u(t)$ are input and output variables, such as
voltage and current, $g(x,u,t)$ is a generalized response
(memresistance $R$, or memductance $G$), $x$ is a n-dimensional
vector describing the internal state of the device, and $f(x,u,t)$
is a continuous n-dimensional vector function \cite{litr7,litr8}.

\begin{figure}[tp]
\centerline{\includegraphics[width=5.5cm]{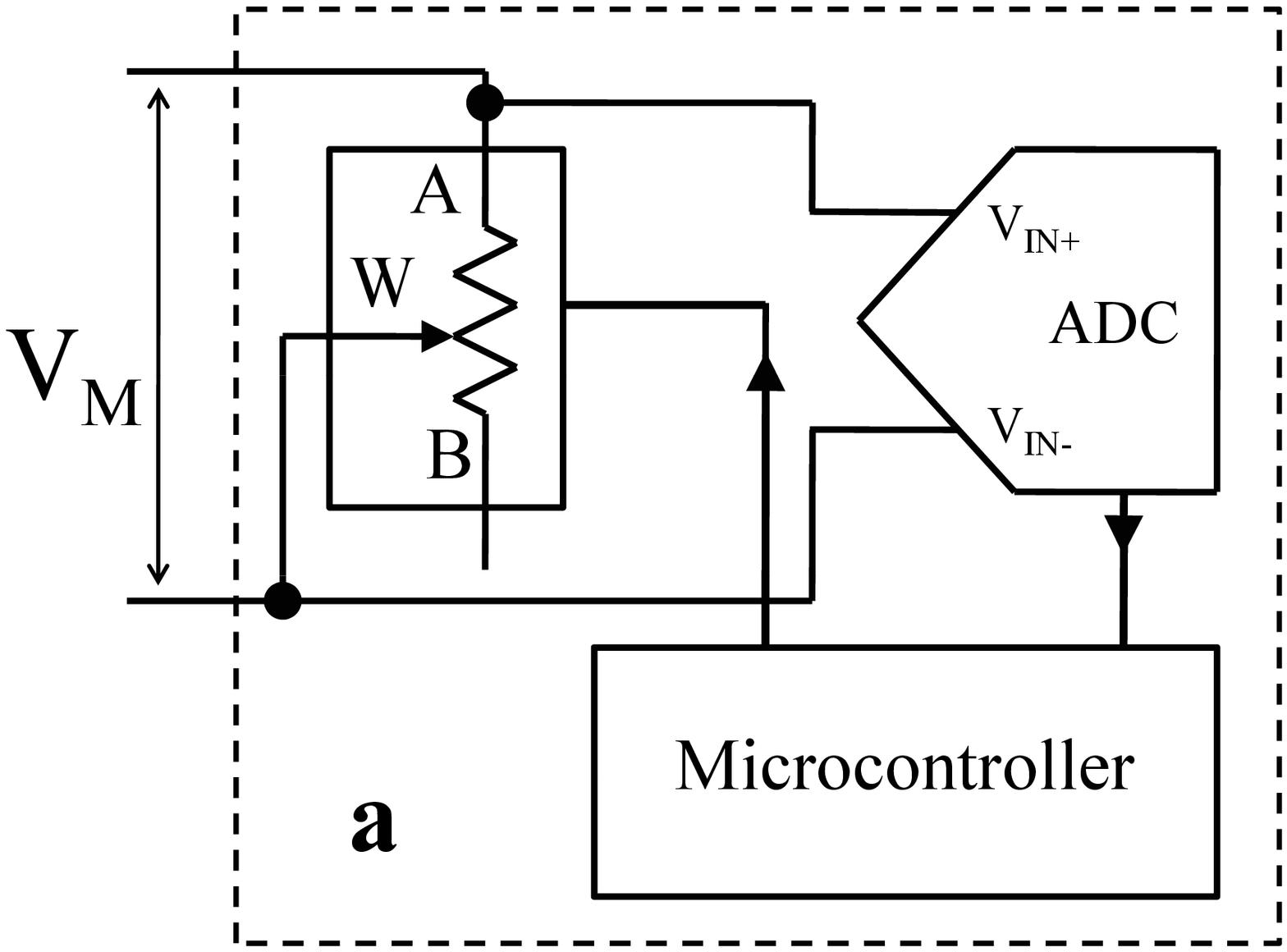}}
\centerline{\includegraphics[width=8cm]{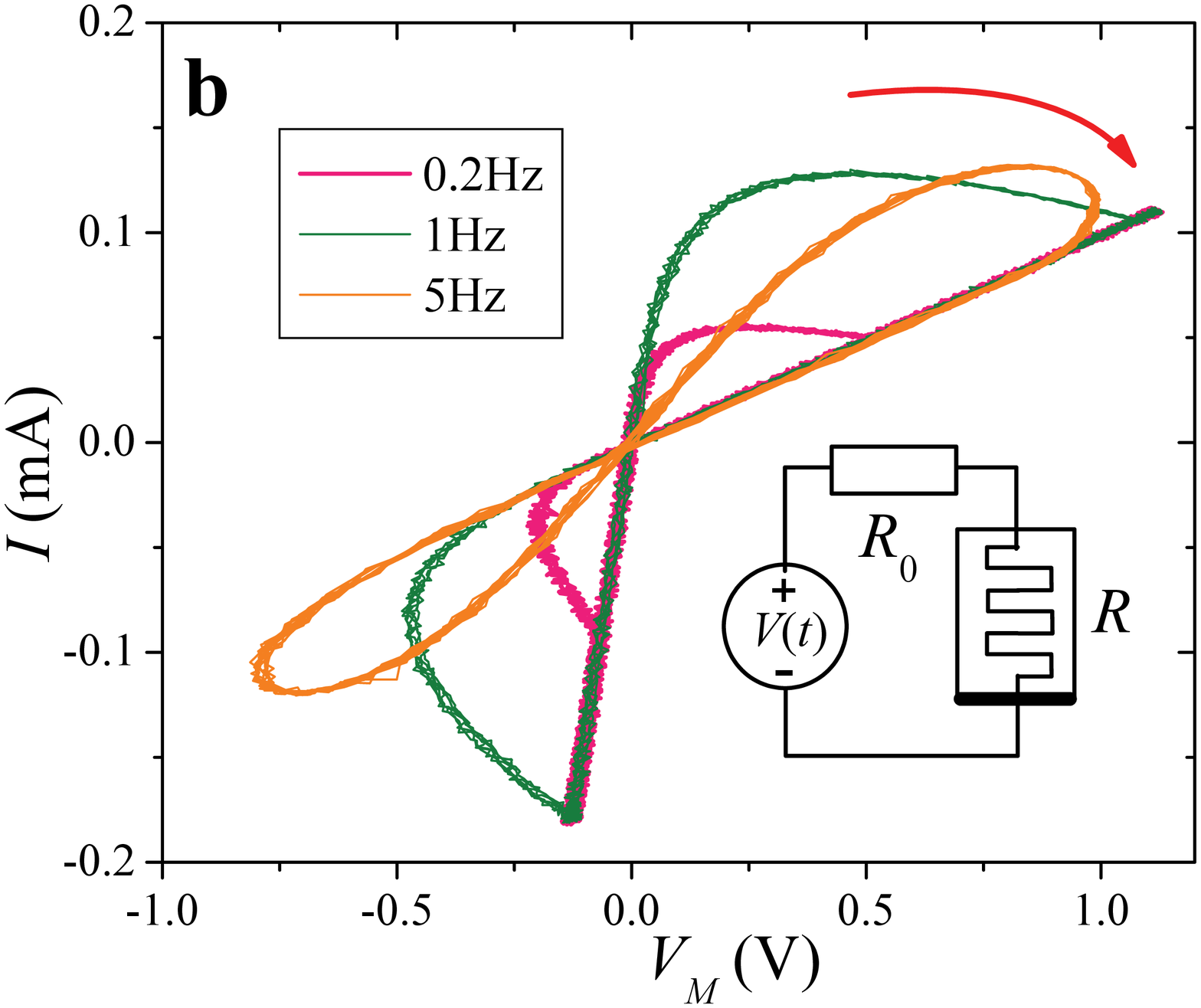}}
 \caption{Electronic synapse. {\bf a}, Schematic of the main units of the
memristor emulator. The memristor emulator consists of a digital
potentiometer, ADC and microcontroller. The digital potentiometer
unit represents an element whose resistance is defined by a
digital code written in it. Two terminals of this unit (A and W)
are the external connection terminals of the memristor emulator.
The differential ADC converts the voltage between A and W
terminals of the digital potentiometer into a digital value. The
microcontroller reads the digital code from ADC and generates (and
writes) a code for the digital potentiometer according to
predefined functions $g(x,u,t)$ and $f(x,u,t)$ and Eqs.
(\ref{eq1}-\ref{eq2}). These operations are performed
continuously. In our circuit, we used a 256 positions 10k$\Omega$
digital potentiometer AD5206 from Analog Device and
microcontroller dsPIC30F2011 from Microchip with internal 12bits
ADC. {\bf b}, Measurements of memristor emulator response when
$V(t)=V_0 \cos (2\pi\omega t)$ with $V_0\simeq 2$V amplitude is
applied to the circuit shown in the inset with $R_0=10$k$\Omega$.
The following parameters determining the memristor emulator
response (see Eqs. (\ref{eq3},\ref{eq4})) were used:
$\alpha=\beta=146$k$\Omega$/(V$\cdot$s), $V_T=4$V, $R_1 =
675\Omega$, $R_2=10$k$\Omega$. We noticed that the initial value
of $R_M$ (in the present case equal to $10$k$\Omega$) does not
affect the long-time limit of the I-V curves. The signals were
recorded using a custom data acquisition system. \label{fig3}}
\end{figure}

As Fig. \ref{fig3}a illustrates schematically, our memristor
emulator consists of the following units: a digital potentiometer,
an analog-to-digital converter and a microcontroller. The A (or B)
terminal and the Wiper of the digital potentiometer serve as the
external connections of the memristor emulator. The resistance of
the digital potentiometer is determined by a code written into it
by the microcontroller. The code is calculated by the
microcontroller according to Eqs. (\ref{eq1}) and (\ref{eq2}). The
analog-to-digital converter provides the value of voltage applied
to the memristor emulator needed for the digital potentiometer
code calculation. The applied voltage can be later converted to
the current since the microcontroller knows the value of the
digital potentiometer resistance.

In our experiments, we implemented a threshold model of
voltage-controlled memristor previously suggested in our earlier
paper \cite{litr16} and also discussed in Ref. \cite{litr8}. In
this model, the following equations (which are a particular case
of Eqs. (\ref{eq1},\ref{eq2})) were used:

\begin{eqnarray}
G&=&x^{-1}, \label{eq3} \\ \dot x&=&\left(\beta V_M+0.5\left(
\alpha-\beta\right)\left[ |V_M+V_T|-|V_M-V_T| \right]\right)
\nonumber\\ & &\times \theta\left( x-R_1\right) \theta\left(
R_2-x\right) \label{eq4},
\end{eqnarray}
where $\theta(\cdot)$ is the step function, $\alpha$  and $\beta$
characterize the rate of memristance change at $|V_M|\leq V_T$ and
$|V_M|> V_T$, respectively, $V_M$ is a voltage drop on memristor,
$V_T$ is a threshold voltage and  $R_1$ and $R_2$ are limiting
values of memristance. In Eq. (\ref{eq4}), the
$\theta$-functions symbolically show that the memristance
can change only between $R_1$ and $R_2$. In the actual software
implementation, the value of $x$  is monitored at each time step
and in the situations when $x<R_1$ or $x>R_2$, it is set equal to
$R_1$ or $R_2$, respectively. In this way, we avoid situations when
$x$ may overshoot the limiting values by some amount and thus not
change any longer because of the step function in Eq.
(\ref{eq4}). In simple words, the memristance changes between
$R_1$ and $R_2$ with different rates $\alpha$ and $\beta$ below
and above the threshold voltage. This activation-type model was
inspired by recent experimental results on thin-film memristors
\cite{litr10} and as we discuss below it reproduces synapse
plasticity.

\begin{figure*}[t]
\begin{minipage}[b]{0.5\linewidth}
\centerline{\includegraphics[width=8cm]{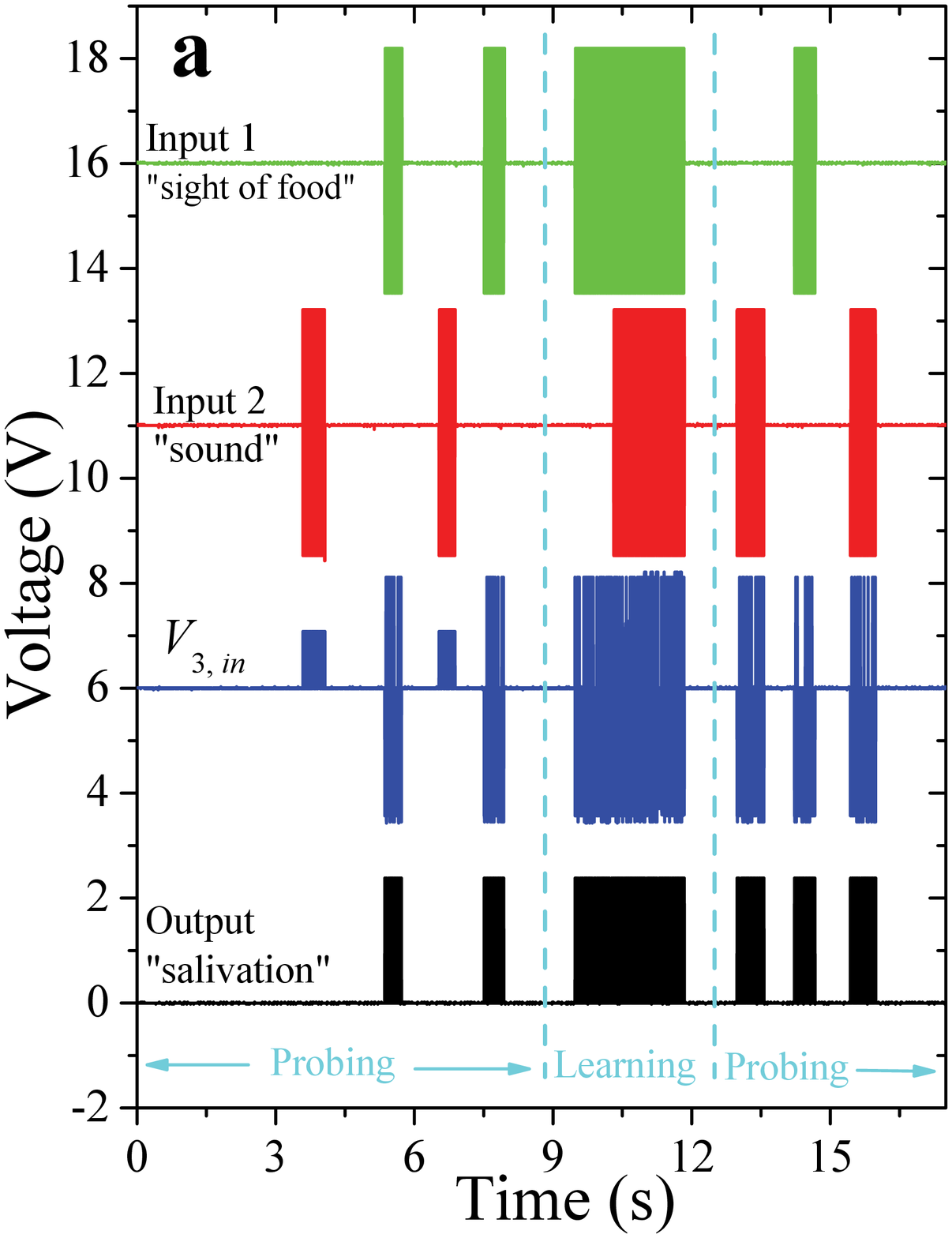}}
\end{minipage}
\begin{minipage}[b]{0.5\linewidth}
\centerline{\includegraphics[width=8cm]{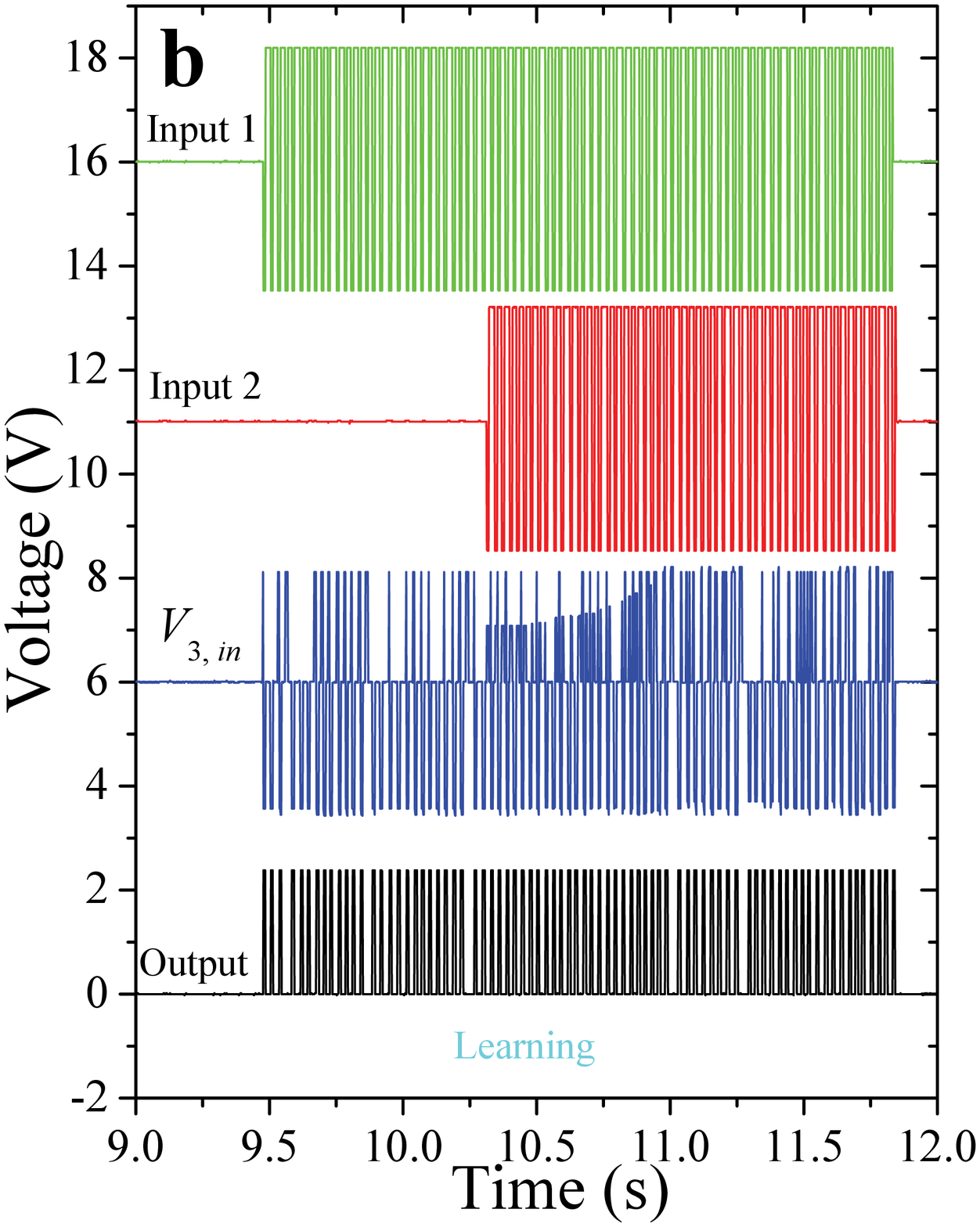}}
\end{minipage}
 \caption{Development of associative memory. {\bf a}, Using electronic neurons
and electronic synapses (memristor emulators), we have built an
electronic circuit corresponding to the neural network shown in
Fig. \ref{fig1}. We have used the following parameters defining
the operation of memristor emulators: $V_T=4$V, $\alpha=0$,
$\beta=15$k$\Omega$/(V$\cdot$s). At the initial moment of time,
the resistance of $S_1$ was selected equal to $R_1=675\Omega$
(lowest resistance state of memristor) and the resistance of $S_2$
was selected equal to $R_2=10$k$\Omega$ (highest resistance state
of memristor). Correspondingly, in the first probing phase, when
Input 1 and Input 2 signals are applied without overlapping, the
output signal develops only when Input 1 signal is applied. In the
learning phase (see also {\bf b} for more detailed picture), Input
1 and Input 2 signals are applied simultaneously. According to
Hebbian rule, simultaneous firing of input neurons leads to
development of a connection, which, in our case, is a transition
of the second synapse $S_2$ from high- to low-resistance state.
This transition is clearly seen as a growth of certain pulses in
the signal $V_{3,in}$ (voltage at the input of third neuron) in
the time interval from 10.25s to 11s in {\bf b}. In the subsequent
probing phase, we observe that "firing" of any input neuron
results in the "firing" of output neuron, and thus an associative
memory realization has been achieved. The curves in {\bf a} and
{\bf b} were displaced for clarity. \label{fig4}}
\end{figure*}

To test that our emulator does indeed behave as a memristor, we
have used the circuit shown in the inset of Fig. \ref{fig3}b, in
which an ac voltage is applied to the memristor emulator, $R$,
connected in series with a resistor $R_0$  which was used to
determine the current. The obtained current-voltage (I-V) curves,
presented in Fig. \ref{fig3}b, demonstrate typical features of
memristive systems. For instance, all curves are pinched
hysteresis loops passing through (0,0) demonstrating no energy
storage property of memristive systems~\cite{litr7,litr81,litr8}.
Moreover, the frequency dependence of the curve is also typical
for memristive systems: the hysteresis shrinks at low frequencies,
when the system has enough time to adjust its state to varying
voltage, and at higher frequencies, when the characteristic
timescale of system variables change is longer than the period of
voltage oscillations.

\subsection{Associative memory}

Using the electronic neurons and electronic synapses described
above, we have built an electronic scheme corresponding to the
neural network depicted in Fig. \ref{fig1}. In this scheme, we
directly connect the memristor terminals to the third neuron
input. In such configuration, our network behaves essentially as a
linear perceptron \cite{perceptr1,perceptr2,perceptr3}, although
different connection schemes are possible. For example, putting a
capacitor between the ground and input of the third neuron, we
would obtain an integrate-and-fire model \cite{spiking_models}. In
such a circuit (and also in perceptron networks with many
synapses) it is important to ensure that the current from a
synapse does not spread to its neighbors. This can be achieved by
placing diodes between the right terminals of synapses in Fig.
\ref{fig1} and the third neuron's input. For our neural network
containing only two synapses the effect of current spreading
between synapses is not important. Moreover, we would like to
highlight that our neural network is fully asynchronous, in
distinction to a scheme suggested by Snider \cite{litr15} based on
a global clock. This makes our approach free
of synchronization issues when scaling up and closer to a bio-inspired circuit. An asynchronous
memristor-based network was also discussed recently
\cite{nature_proc}.

Fig. \ref{fig4} demonstrates the associative memory development in
the present network. Our experiment consists in application of stimulus
signals to the first ("sight of food") and second ("sound")
neurons, and monitoring of the output signal on the third
("salivation") neuron. We start from a state when the first
synaptic connection is strong (low resistance state of the first
memristor) and second synaptic connection is weak (high resistance
state of the second memristor).

In the first "probing phase" ($t<9$s, Fig. \ref{fig4}) we apply
separate non-overlapping stimulus signals to the "sight of food"
and "sound" neurons. This results in the "salivation" neuron
firing when a stimulus signal is applied to the "sight of food"
neuron, but not firing when a stimulus signal is applied to the
"sound" neuron. Electronically, it occurs because pulses generated
by the "sight of food" neuron exceed the threshold voltage of the
"salivation" neuron (due to a low resistance of the first
memristor synapse) while the voltage on the "salivation" neuron
input due to the "sound" neuron pulses is below the threshold
voltage of the "salivation" neuron. In this phase there is no
memristor state change since the first memristor is already in its
limiting state (with minimal resistance allowed) and its
resistance cannot decrease further, and voltage drop on the second
memristor is below its voltage threshold.

In the "learning phase" (9s$<t<$12s, Fig. \ref{fig4}), stimulus
voltages are applied simultaneously to both input neurons, thus
generating trains of pulses. The pulses from different neurons are
uncorrelated, but sometimes they do overlap, owing to a random
component in the pulse separation (see Fig. \ref{fig2} caption for
details). During this phase, in some moments of time,
back-propagating pulses from the "salivation" neuron (due to
excitation from the "sight of food" neuron) overlap with forward
propagating pulses from the "sound" neuron causing a high voltage
across the second memristor synapse. As this voltage exceeds the
memristor threshold, the second synapse state changes and it
switches into a low resistance state. It is important to note that
this change is possible when both stimuli are applied together (in
other words they correlate). As a result, an association between
input stimuli develops and the circuit "learns" to associate the
"sight of food" signal to the "sound" signal.

Our measurements during the second probing phase ($t>12$s, Fig.
\ref{fig4}) clearly demonstrate the developed association. It is
obvious that, in this phase, any type of stimulus - whether from the
"sight of food" or from the "sound" neurons - results in the
"salivation" neuron firing.

Mention should be made about the memristor evolution function
given by Eq. \ref{eq4}. In the present neural network the
resistance of the second synapse can only decrease, since only
negative or zero voltages can be applied to the memristor. A process
of synapse depression (corresponding to an increased synapse resistance
in our electronic network) can be easily taken into account in different
ways. The simplest way is to add a small positive
constant to the first line of Eq. \ref{eq4} (in the parentheses)
for the conditioned memristor (S$_2$ in Fig.~\ref{fig1}). Then, even at zero value of
$V_M$, the resistance of the memristor will slowly increase.
Another way is to apply a small positive bias to the conditioned
memristor (S$_2$) in the circuit without any changes in Eq.
\ref{eq4} (this would also require taking a non-zero $\alpha$).

\section{Conclusion}
We have shown that the electronic (memristive) synapses and
neurons we have built can represent important functionalities of
their biological counterparts, and when combined together in
networks - specifically the one represented in Fig. \ref{fig1} of
this work - they give rise to an important function of the brain,
namely associative memory. It is worth again mentioning that,
although other memristors (e.g., those built from oxide thin films
\cite{litr17}) could replace the emulator we have built, the
electronic neurons and synapses proposed here are electronic
schemes that can be built from off-the-shelf inexpensive
components. Together with their extreme flexibility in
representing essentially any programmable set of operations, they
are ideal to study much more complex neural networks \cite{litr18}
that could adapt to incoming signals and "take decisions" based on
correlations between different memories. This opens up a whole new set of
possibilities in reproducing different types of learning and
possibly even more complex neural processes.

\section*{Acknowledgment}
M.D. acknowledges partial support from the National Science
Foundation (DMR-0802830).


\begin{thebibliography}{1}

\bibitem{litr1} J.~R. Anderson, {\it Language, memory, and thought}. Hillsdale, NJ:
Erlbaum, 1976.

\bibitem{litr2} I.~P. Pavlov, {\it Conditioned Reflexes: An Investigation of the
Physiological Activity of the Cerebral Cortex} (translated by G.
V. Anrep). London: Oxford University Press, 1927.

\bibitem{hopf} J.~J. Hopfield, "Neural networks and physical systems with emergent collective computational properties,"
{\it Proc. Nat. Acad. Sci. (USA)}, vol. 79, pp. 2554-2558, 1982.

\bibitem{litr3} K.~Gurney, {\it An Introduction to Neural Networks}. UCL Press
(Taylor \& Francis group), 1997.

\bibitem{litr5} L.~N. Cooper, "Memories and memory: a physicist's approach to
the brain," {\it Int. J. Modern Physics A}, vol. 15, pp.
4069-4082, 2000.

\bibitem{litr4} T.~Munakata,  {\it Fundamentals of the New Artificial Intelligence:
Neural, Evolutionary, Fuzzy and More}, 2nd ed. Springer, 2008.

\bibitem{lewis} N.~Lewis and S.~Renaud, "Spiking neural networks "in silico": from single neurons to large scale networks",
Systems, Signal and Devices Conference, SSD 2007, Hammamet, Tunisia,
March 19-22 2007.

\bibitem{widrow} B. Widrow, ``Rate of adaptation of control
systems'', ARS Journal, pp. 1378-1385 (1962).

\bibitem{no_plast1} M.~Mahowald, and R.~Douglas, "A silicon neuron", {\it Nature}, vol.354,
pp. 515-518, 1991.

\bibitem{no_plast2} R.~Jung, E.~J. Brauer, and J.!J. Abbas "Real-time interaction between
a neuromorphic electronic circuit and the spinal cord", {\it  IEEE
Trans. on Neural Systems and Rehab. Eng.}, vol. 9, pp.319–326,
2001.

\bibitem{no_plast3} G.~LeMasson, S.~Renaud, D.~Debay, and T.~Bal, "Feedback
inhibition controls spike transfer in hybrid thalamic circuits",
{\it Nature}, vol. 4178, pp. 854-858, 2002.

\bibitem{no_plast4} M.~Sorensen, S.~DeWeerth, G.~Cymbalyuk, R.~L. Calabrese,
"Using a hybrid neural system to reveal regulation of neuronal
network activity by an intrinsic current", {\it J. Neurosci.},
vol.24, pp. 5427-5438, 2004.

\bibitem{plast1} R.J. Vogelstein, U. Malik, G. Cauwenberghs, "Silicon spikebased
synaptic array and address-event transceiver", Proceedings of
ISCAS'04, vol.5, pp.385-388, 2004.

\bibitem{plast2} B. Glackin, T.M. McGinnity, L.P. Maguire, QX Wu, A.
Belatreche, "A novel approach for the implementation of large
scale spiking neural networks on FPGA hardware", Proc. IWANN 2005
Computational Intellingence and Bioinspired Systems, pp.552-563,
Barcelona, Spain, June 2005.

\bibitem{plast3} J. Schemmel, K. Meier, E. Mueller, "A new VLSI model of neural
microcircuits including spike time dependent plasticity", Proc.
ESANN'2004, pp. 405-410, 2004.

\bibitem{plast4}G.~Indiveri, E.~Chicca, and R.~Douglas. "A VLSI array of low-power spiking neurons and bistable synapses with spike-timing dependent
plasticity", {\it IEEE Transactions on Neural Networks}, vol. 17,
pp. 211-221, 2006.

\bibitem{plast5} J.~V. Arthur and K. Boahen, "Learning in Silicon: Timing is Everything",
{\it Advances in Neural Information Processing Systems}, vol. 18,
Eds. B.~Sholkopf and Y.~Weiss,  MIT Press, 2006.

\bibitem{plast6} A.~Bofill and A.~F. Murray. "Circuits for VLSI implementation of temporally
    asymmetric Hebbian learning", {\it Advances in Neural Information processing systems}, vol. 14,
    Eds. T.~G. Dietterich, S.~Becker, and Z.~Ghahramani,
    editors,  MIT
    Press, Cambridge, MA, 2001.

\bibitem{plast7} B.~Linares-Barranco, E.~Sánchez-Sinencio, A.~Rodríguez-Vázquez, and J.~L.
Huertas, "A CMOS Analog Adaptive BAM with On-Chip Learning and
Weight Refreshing", {\it IEEE Trans. Neural Networks}, vol. 4, pp.
445-455, 1993.

\bibitem{litr6} L.~O. Chua, "Memristor - The Missing Circuit Element," {\it IEEE Trans. Circuit Theory}, vol. 18, pp.
507-519, 1971.

\bibitem{litr7} L.~O. Chua and S.~M. Kang, "Memrisive devices and systems," {\it Proc. IEEE}, vol. 64, pp. 209-223, 1976.

\bibitem{litr9} D.~B. Strukov, G.~S. Snider, D.~R. Stewart, and R.~S. Williams, "The missing memristor found," {\it Nature (London)}, vol. 453, pp.
80-83, 2008.

\bibitem{litr10} J.~J. Yang, M.~D. Pickett, X.~Li, D.~A.~A. Ohlberg, D.~R.
Stewart, and R.~S. Williams, "Memristive switching mechanism for
metal/oxide/metal nanodevices," {\it Nature Nanotechnology}, vol.
3, pp. 429-433, 2008.

\bibitem{litr81}  M.~Di Ventra, Yu.~V. Pershin, and L.~Chua, "Putting memory into circuit elements: memristors, memcapacitors and meminductors,"
{\it Proc. IEEE}, vol. 97, pp. 1371-1372, 2009.

\bibitem{litr8}  M.~Di Ventra, Yu.~V. Pershin, and L.~Chua, "Circuit elements with memory: memristors, memcapacitors and meminductors,"
{\it Proc. IEEE}, vol. 97, pp.  1717-1724, 2009.

\bibitem{litr11} T.~Driscoll, H.-T. Kim, B.-G. Chae, M.~Di Ventra, and D.~N. Basov, "Phase-transition driven memristive system," {\it Appl. Phys. Lett.} vol. 95, pp. 043503/1-3, 2009.

\bibitem{litr12} Yu.~V. Pershin and M.~Di Ventra, "Spin memristive systems: Spin memory effects in semiconductor spintronics,"
{\it Phys. Rev. B, Condens. Matter}, vol. 78, pp. 113309/1-4, 2008.

\bibitem{litr13} G.~Stuart, M.~Spruston,  B.~Sakmann, and M.~Häusser, "Action
potential initiation and backpropagation in neurons of the
mammalian CNS," {\it Trends Neurosci.}, vol. 20, pp. 125-131,
1997.

\bibitem{litr14} J.~Waters, A.~Schaefer, and B.~Sakmann, "Backpropagating action
potentials in neurons: measurement, mechanisms and potential
functions," {\it Prog. Biophys. Mol. Biol.}, vol. 87, pp. 145-170,
2005.

\bibitem{litr15} G.~S. Snider, "Cortical Computing with Memristive
Nanodevices," {\it SciDAC Review}, vol. 10, pp. 58-65, 2008.

\bibitem{litr16}  Yu.~V. Pershin, S.~La Fontaine and M.~Di
Ventra, "Memristive model of amoeba's learning," {\it Phys. Rev.
E}, vol. 80, pp. 021926/1-6, 2009.

\bibitem{perceptr1} R.~Rojas, "Neural Networks - A Systematic
Introduction" (Springer-Verlag, Berlin), 1996.

\bibitem{perceptr2} F. Rosenblatt, "The Perceptron: A Probabilistic Model for
Information Storage and Organization in the Brain",  {\it Psych.
Rev.}, vol. 65, pp. 386-408, 1958.

\bibitem{perceptr3} M. L. Minsky and S. A. Papert, "Perceptrons" (Cambridge, MA: MIT
Press), 1969.

\bibitem{spiking_models} W. Gerstner and W. Kistler, "Spiking neuron models" (Cambridge
university press), 2002.

\bibitem{nature_proc} B. Linares-Barranco1 and T.
Serrano-Gotarredona, "Memristance can explain
Spike-Time-Dependent-Plasticity in Neural Synapses",
http://hdl.handle.net/10101/npre.2009.3010.1,  2009.

\bibitem{litr17} J.~Borghetti, Z.~Li, J.~Straznicky, X.~Li, D.~A.~A. Ohlberg, W.~Wu, D.~R. Stewart, and R.~S. Williams, "A hybrid
nanomemristor/transistor logic circuit capable of
self-programming," {\it PNAS}, vol. 106, pp. 1699-1703, 2009.

\bibitem{litr18} K.~Likharev, A.~Mayr, I.~Muckra, and O.~Turel, "CrossNets:
High-Performance Neuromorphic Architectures for CMOL Circuits,"
{\it Ann. NY Acad. Sci.}, vol. 1006, pp. 146-163, 2003.

\end{thebibliography}
\end{document}